\documentclass[conference,letterpaper]{IEEEtran}

\usepackage[utf8]{inputenc}
\usepackage{amsmath, amssymb, amsthm}
\usepackage{enumerate}
\usepackage{graphicx}
\usepackage{hyperref}
\usepackage{xcolor}
\usepackage{booktabs}
\usepackage[utf8]{inputenc} 
\usepackage[T1]{fontenc}
\usepackage{url}   
\usepackage[noadjust]{cite}
\usepackage {tikz}
\usepackage{pgfplots}
\pgfplotsset{compat=newest,compat/show suggested version=false}
\usetikzlibrary {shapes,arrows,positioning}
\usetikzlibrary {arrows.meta} 
\usepackage{multirow}
\usepackage{color, colortbl}
\usepackage{algorithm, algorithmicx, algpseudocode}
\usepackage{bm}
\usepackage{mathrsfs}
\usepackage{mathtools}
\usepackage{lipsum}

\usepackage[format=plain]{subcaption}
\DeclareMathAlphabet{\mathbfsl}{OT1}{ppl}{b}{it} 

\DeclareMathOperator*{\argmax}{arg\,max}

\newtheoremstyle{definitionstyle}
  {\topsep}   
  {\topsep}   
  {\normalfont}  
  {}          
  {\bfseries} 
  {}         
  {.5em}      
  {\thmname{#1}\thmnumber{ #2}\thmnote{ [\normalfont#3]}} 

\theoremstyle{definitionstyle}
\newtheorem{definition}{Definition}

\theoremstyle{definition}

\theoremstyle{remark}
\newtheorem{remark}{Remark}

\begin{document}

\hypersetup{
    hidelinks
}


\title{Efficient Decoders for Sensing Subspace Code}

\IEEEoverridecommandlockouts
\author{%
  \IEEEauthorblockN{Siva Aditya Gooty and
Hessam Mahdavifar 
}
  \IEEEauthorblockA{
  Department of
Electrical and Computer Engineering, Northeastern University, Boston, MA
02115, USA.}
Email: \{\href{mailto:gooty.s@northeastern.edu}{gooty.s}, \href{mailto:h.mahdavifar@northeastern.edu}{h.mahdavifar}\}@northeastern.edu

                    
\thanks{This work was supported by NSF under Grant CCF-2415440 and the Center for Ubiquitous Connectivity (CUbiC) under the JUMP 2.0 program.}
 %
 }
\maketitle


    


\begin{abstract}

Sparse antenna array sensing of source/target via direction of arrival (DoA) estimation motivates design of the sensing framework in joint communication and sensing (JCAS) systems for sixth generation (6G) communication systems. Recently, it is established by Mahdavifar, Rajam\"{a}ki, and Pal that array geometry of sparse arrays has fundamental connections with the design of subspace codes in coding theory. This was then utilized to design efficient \textit{sensing subspace codes} that estimate the DoA with good resolution. Specifically, the Bose-Chowla sensing subspace code provides near optimal code design for unique DoA estimation with tight theoretical upper bound on the error performance. However, the currently known decoder for these codes, to estimate the DoA, is a traditional \textit{Maximum-a-Posterior (MAP) decoder} with complexity that is cubic with the number of antennas. In this work, we propose novel efficient decoding algorithms for sensing subspace codes, that reduce the complexity down to quadratic while providing new knobs to tune in order to tradeoff complexity with error performance. The decoders are further evaluated for their performance via Monte Carlo simulations for a range of SNRs demonstrating promising performance that smoothly approaches the MAP performance as the complexity grows from quadratic to cubic in the number of antennas.

\end{abstract}


\section{Introduction}
\label{sec:introduction}
Antenna array system design has been one of the major driving factor for upcoming 6G wireless communication innovation and design~\cite{Guo_Antenna, Wang_2025}. New multi-antenna structures that facilitate multiple-input multiple-output (MIMO) communication govern efficient design of physical layer communication systems~\cite{Wang_XLMIMO}. With MIMO systems predominantly enabling the capacity and bandwidth intensive physical layer design with novel beamforming techniques~\cite{Dreifuerst_XLMIMO}, the antenna array design has been one of the key research directions for 6G.

Recently, the design focus is shifting from traditional dense antenna array systems consisting of uniform array to sparse antenna arrays for various advantages such as, lower power consumption, better resolution for classical DoA estimation problem, reduced hardware cost, etc~\cite{Vaidyanathan_Sparse, Roberts_Sparse}. The DoA estimation framework is extensively studied to characterize the JCAS research in 6G and beyond with smart antenna array designs aiding pragmatic sensing strategies~\cite{Rajamaki_SS,Gooty_JCAS}. 

The sparse antenna \textit{array geometry} configurations provide interesting connections to sensing strategies which ultimately result in better DoA estimation frameworks further aiding in improved communication capacities~\cite{Vaidyanathan_Sparse}. The physical positioning of the sparse antennas for passive sensing has a pivotal relation to subspace codes, with Mahdavifar et al. in~\cite{Mahdavifar_SS} providing the first of its kind, a promising connection between sparse array geometry and subspace codes. The subspace codes have numerous advantages with applications including but not limited to non-coherent networks, beamforming codebook design for MIMO systems, etc~\cite{Soleymani_CP, gooty2025precodingdesignlimitedfeedbackmiso}. The new connection established between sparse antenna array with subspace coding has given rise to \textit{sensing subspace codes}, which produce efficient designs for DoA estimation of source/target. This novel subspace code construction for sparse arrays was tested for single source DoA estimation with Bose-Chowla constructions that provide near-optimal \textit{minimum subspace distance}. The Bose-Chowla sensing subspace codebook with its minimum subspace distance between the codewords provide unique mapping to the DoA resolution. In addition, the Bose-Chowla sensing subspace code offers a significantly lower probability of error as compared to traditional uniform linear array design. Theoretical bounds characterized the robustness of the code with unique relation to the codebook size with respect to the array factor. However, the decoder employed is a traditional \textit{MAP decoder}, to estimate the DoA with theoretical time complexity growing at cubic time with respect to the number of active sparse antenna elements. 

In this paper, we propose efficient decoding algorithms for Bose-Chowla sensing subspace codes of~\cite{Mahdavifar_SS}, that scale quadratically with number of antenna elements. We first propose a \textit{window decoder} that provides lower run-time complexity compared to the MAP decoder, with a trade-off between error performance and decoding complexity by tuning the underlying window size. Furthermore, we propose an array geometry based decoding algorithm that provides flexibility at the receiver to choose only a subset of antennas in the Bose-Chowla-based geometry with certain number-theoretic properties that enable faster quadratic decoding with a tradeoff between error performance and time complexity. Specifically, three array geometry based decoders are presented-\textit{geometric, modified geometric and geometric-reduced MAP}. The geometric decoder is the simplest of the all with modified geometric decoder being the next followed by geometric-reduced MAP. The time complexity of all these decoders can be limited to the order of quadratic time, by flexibly choosing the antennas to decode the received signal to estimate the DoA. The performance of all the decoders are evaluated for given values of SNR via Monte Carlo simulations demonstrating promising performance that smoothly approaches the MAP performance as the complexity grows from quadratic to cubic, in the number of antennas.


\section{Preliminaries: Sensing Subspace Codes}
\label{sec:Preliminaries}
\subsection{Notation Convention}
\label{ssec:Notation}

For $N \in \mathbb{N}$, the set of $\{1,2,\ldots,N\}$ is denoted by $[N]$. The set of $N$ roots of unity is denoted by $\zeta(N) \overset{\text{def}}{=} \{e^{j2\pi n/N}, n \in [N]-1\}$. The matrix and its column space is denoted by $\mathbf{X}$ and $\langle \mathbf{X} \rangle$ respectively. For a given vector space $\mathcal{W}$, the set $\mathcal{P(W)}$, denotes the set of subspaces of $\mathcal{W}$. Let $\mathcal{W} = \mathbb{L}^M$, where $\mathbb{L}$ can be either $\mathbb{C}$ or $\mathbb{R}$. The set of all $t$ dimensional subspaces of $\mathcal{W} = \mathbb{L}^M$ is referred to as Grassmannian space, denoted by $G_{t, M}(\mathbb{L}) \overset{\text{def}}{=} \{\langle \mathbf{X} \rangle: \mathbf{X} \in \mathbb{C}^{M \times t}, \mathbf{X}^{\rm H}\mathbf{X} = \mathbf{I}_{t} \}$, where $\mathbf{I}_t$ is an $t \times t$ identity matrix. 

\subsection{Subspace Codes}
\label{ssec:BCG codes}
 
Consider two $t$ dimensional subspaces, $\mathbf{U}$ and $\mathbf{V}$. Let $\mathbf{u}_i \in \mathbf{U}$ and $\mathbf{v}_i \in \mathbf{V}$ be column vectors with unit length such that $|{\mathbf{u}_i}^{\rm H}{\mathbf{v}_i}|$ is maximal, subject to ${\mathbf{u}_i}^{\rm H} \mathbf{u}_j = 0 $ and ${\mathbf{v}_i}^{\rm H} \mathbf{v}_j = 0$, for all $i \neq j $ and $j>i\geq 1$.  Then the \textit{principal} angle $\beta_i$, for $i \in [t]$ between $\mathbf{U}$ and $\mathbf{V}$ is 
    $\beta_i= \arccos|\mathbf{v}^{\rm H}_i\mathbf{u}_j|$,
see, e.g~\cite{Barg_Grassmann, Roy_principal, Soleymani_CP, Mahdavifar_SS}. 
The subspace distance between any two subspaces is defined via \textit{quasi-distance} metric~\cite{Mahdavifar_SS} as, 
\begin{align}
\label{eqn:quasi_distance}
    d^{(s)}(\mathbf{U},\mathbf{V}) = \sum_{i=1}^{t}\sin^2(\beta_i).
\end{align}
Note that~\eqref{eqn:quasi_distance} is proportional to the square of the chordal distance metric, $d_c$~\cite{Mahdavifar_SS,Soleymani_CP, Conway1996}.
For any analog subspace code $\mathcal{C} \in \mathcal{P(W)}$,
the minimum subspace distance is given by,
\begin{align}
\label{eqn: min subspace distance}
    d^{(s)}_{\text{min}}(\mathcal{C}) \overset{\text{def}}{=} \min_{\mathbf{U,V} \in \mathcal{C}, \mathbf{U \neq V}} d^{(s)}(\mathbf{U,V}).
\end{align} 
Sensing subspace codes were introduced in \cite{Mahdavifar_SS} (see, [\cite{Mahdavifar_SS}, Definition~1]), establishing formal connections between chordal distance and array beamforming pattern through the selection of antenna geometry.

\subsection{System Model}
\label{ssec: System Model}
The $M$ antenna receiver \textit{passive sensing} system with single time sample is designed to estimate the DoA $\theta$ of a single source/target in the far field~\cite{Mahdavifar_SS}. We consider one dimensional Bose-Chowla sensing subspace code with near-optimal design for single source/target \textit{spatial sensing}. Formally, the definition for Bose-Chowla sensing subspace code, 
is presented next.
\begin{definition}[Bose-Chowla sensing subspace code ($\mathcal{C^{(BC)}})$ ~\cite{Mahdavifar_SS}]
For a \textit{good} choice of difference set $\mathcal{A} = \{d_1, d_2, \ldots, d_M\}$ given by Bose-Chowla construction~\cite{Chowla1962_63}, which is a construction for Golomb ruler \cite{Sidon1932}, $\mathcal{C^{(BC)}}$ is given by,
\begin{align}
    \mathcal{C^{(BC)}} = \{(\langle \alpha^{d_1},\alpha^{d_2},
\ldots, \alpha^{d_M}\rangle): \alpha \in \zeta(N)\}
\end{align}
where $\alpha = e^{j\pi\sin\theta}$ and $|\mathcal{C}| = N$.
\end{definition}

Along the direction $\theta$, when the single source/target transmits information to the receiver, the planar wave impinges on the receiver to generate the array manifold vector $\mathbf{h}(\theta) = [h_m(\theta)]_{M \times 1}, m \in [M]$ with relative phase shifts experienced by the antennas. At any $m^{th}$ antenna, the channel gain is, 
\begin{align}
    h_m(\theta) = e^{j\pi d_m \sin\theta}, \text{for } m \in [M].
\end{align}
With array geometry $\{d_m\}_{m=1}^{M}$ fixed by $\mathcal{C^{(BC)}}$, the receiver is tasked with estimating the DoA $\theta$. The DoA $\theta$ is assumed to originate from $N$ unique grid points with $\sin(.)$ of the grid points equally distanced in $[-1,1)$. Specifically, $\theta \in \{\arcsin(-1+2(n-1)/N)\}_{n=1}^{N}$. For $\mathcal{C^{(BC)}}$, the alphabet size of $\theta$ is $N=M^2 - 1$. Note that for a given $\alpha \in \zeta(N)$, there exists a codeword $\mathbf{c}(\alpha) \in \mathcal{C}$ such that
\begin{align}
    {\mathbf{c}}(\alpha) = (\langle \alpha^{d_1},\alpha^{d_2},
\ldots, \alpha^{d_M}\rangle).
\end{align}
This also implies that there exists a unique mapping from array manifold vector to a corresponding codeword $\mathcal{\mathbf{c}}(\alpha)\in \mathcal{C^{(BC)}}$, see \textit{DoA sensing encoder} [\cite{Mahdavifar_SS}, Definition~1].

The signal received from the receiver $\mathbf{y}_{M\times 1}$ is of the form
\begin{align}
    \mathbf{y}_{M\times1} =  \mathbf{h}(\theta)x + \mathbf{w}_{M \times 1}
\end{align}
where $x$ is the unknown transmitted information and $\mathbf{w}$ is the additive noise vector with complex i.i.d. Gaussian entries. For simplicity, $|x| = 1$. The ultimate task of the receiver is to estimate the DoA $\theta$ given $\mathcal{C^{(BC)}}$ and the received $\mathbf{y}$ irrespective of $x$. In the next section, we introduce novel efficient decoders to map the received $\mathbb{C}^M$ signal to estimate the DoA $\theta$.  

\section{Efficient Decoding Algorithms}
\label{sec:Concept}
Before designing any decoder for a new code, the natural choice to begin with is a MAP decoder, which is also a minimum distance decoder with uniform prior, denoted by $\mathcal{D}_\text{min}$. 
In~\cite{Mahdavifar_SS}, the MAP decoder
is 
analyzed with an upper bound on probability of error for different SNR regimes, see [\cite{Mahdavifar_SS}, Corollary~3]. However, the MAP decoder has a cubic complexity in the number of antennas which may not be appealing in practice for large $M$'s. In the remainder of this section, we introduce several novel decoding methods to 
lower the complexity of the decoder. 

\subsection{Window Decoder}
\label{ssec:Window Decodoer}
We fix a window of size $z$ and initialize an empty codebook of size $M \times N/z$ in step 1. In steps 2-4, we add $z$ codewords of $\mathcal{C^{(BC)}}$ together to form a new codebook $\mathcal{C'^{(BC)}}$ of size $N/z$ in each of the $N/z$ windows. The new decoder now spans over $N/z$ combined codewords at step 5, which are the respective $z$ sums of initial $N$ codewords with a hope that any $z$-Window has candidate codeword carrying the DoA $\theta$ information. 
In step 6, set $S$ contains indices of codewords within a candidate window using which the decoder spans within this candidate window at step 7 to declare $\hat{\theta}$ in~\autoref{alg:Win Decoder}. 
\begin{algorithm}
\caption{{Window decoder}}
\begin{algorithmic}[1] 
\State $\mathcal{C'^{BC}} =\text{zeros}(M,N/z)$
\For{$i = 1$ to $N/z$}
    \State $\mathcal{C'^{BC}}[:, i] = \text{sum}\left(\mathcal{C^{BC}}[:, (i-1)*z+1:i*z]\right)$ 
\EndFor
\State $v \gets \argmax\limits_{\forall \mathbf{c} \in \mathcal{C'^{BC}}} \mathbf{y}^{\rm H} * \mathbf{c}$
\State Define the set $S = \{q:(v-1)*z+1\leq q \leq v*z\}$
\State $\hat{\theta} \gets \argmax\limits_{\forall s \in S} (\mathbf{y^{\rm H}}*\mathcal{C^{BC}}[:, s])$
\end{algorithmic}
\label{alg:Win Decoder}
\end{algorithm}

Although theoretical time complexity is $\mathcal{O}(M^3)$, the empirical run-time complexity is now significantly reduced for large values of $M$. In~\autoref{sec:Simulation}, the {window decoder} is empirically tested for different values of window size $z$ and the performance is evaluated along with the MAP decoder.

\subsection{Geometric Decoder}
\label{ssec:GCD Decoding}
We utilize the array geometry of the antenna to effectively quantize the signal across each antenna in~\autoref{alg:Geometric}. The key factor to decode the DoA $\theta$ is to precalculate the greatest common divisor (gcd) of $\{d_m\}_{m=1}^M$ with respect to $N$ grid points. The array geometry which represents antenna positions are already known at the receiver. 
The receiver quantizes the per antenna signal, $\mathbf{y}(m)$, to its closest point on unit circle. 

\begin{algorithm}

\caption{Geometric decoder}
\begin{algorithmic}[1] 
\State $L\gets \emptyset$
\State $\bm{r} \gets \text{gcd}(\mathcal{A}, N)$
\For{$m=1$ to $M$}
\If{$\bm{r}(m)=1$}
    \State $t_m \gets\lfloor (\arg(\mathbf{y}(m)) *(N/2\pi) + 1/2)\rfloor$ 
    \State $\theta_m \gets (\mathcal{A}(m)^{-1}*t_m) \text{ mod } N$
    \State $L \gets \{L, \theta_m\}$
\Else
    \State $t_m \gets\lfloor (\arg(\mathbf{y}(m)) *(N/(\bm{r}(m) *2\pi)) + 1/2)\rfloor$ 
    \State $\bm{t_m'} \gets \text{mod}((0:N-1)*d_m/\bm{r}(m), N/\bm{r}(m))$
    \State $\bm{\theta_m} \gets \text{find}(t_m=\bm{\bm{t'_m}})$
    \State $L \gets \{L, \bm{\theta_m}\}$
\EndIf
\EndFor
\State $\hat{\theta} \gets \text{mode}(L)$

\end{algorithmic}
\label{alg:Geometric}
\end{algorithm}
\vspace{-2mm}
In step 1, we maintain an empty set $L$. In step 2, the gcd values $\bm{r}$ associated with $\mathcal{A}$ is pre-calculated at the receiver. If the value of $\bm{r}(m)=1$ in step 4, inverse of $d_m$ exists. Once the signal received at the antenna is quantized to a point $t_m$ on the unit circle with $N$ roots of unity via step 5, the angle corresponding to this $m^{th}$ antenna will be stored in a set $L$ via step 7. On the other hand, if $\bm{r}(m)\neq1$, the inverse of $d_m$ doesn't exist. This essentially means, there exist multiple possibilities for $\theta_m$. Nevertheless we continue to quantize the signal $\mathbf{y}(m)$ to a unit circle with smaller roots of unity, $N/\bm{r}(m)$ in step 9. In step 10, we consider all $t_m\in[N-1]\cup\{0\}$ points over the smaller unit circle with $N/\bm{r}(m)$ roots of unity as $\bm{\bm{t'_m}}$. Since any point $t_m\in[N-1]\cup\{0\}$ can be quantized $\bm{r}(m)$ times on a smaller unit circle with $N/\bm{r}(m)$ points of unity, we find all $\bm{r}(m)$ number of possible points over a bigger unit circle with $N$ points of unity at step 11. Note that, $\bm{\theta_m}$ in step 11 has $\bm{r}(m)$ values as against a unique value in step 5. In step 12, we append $\bm{\theta_m}$ to the set $L$. Finally in step 15, we declare the most repeated value of $\theta_m$ across $[M]$ as our candidate DoA $\theta$. Step 5, 6, 7 and 9 need constant $\mathcal{O}(1)$ time. Step 7 and 12 can be updated with a hash-map for better storage complexity needs. The time complexity remains at $\mathcal{O}(M^2)$, up to constant factors. Step 10 and 11 can be completed in $\mathcal{O}(M^2)$ time, with step 12 at $\mathcal{O}(\bm{r}(m))$. The entire algorithm can be implemented at the expense of $\mathcal{O}(M^2)$ complexity.  
\begin{remark}
    The overall complexity of the algorithm critically depends 
    $\bm{r}(m)$ for each position of the antenna. Depending on the value $M$, the receiver can choose a subset of $M$ antennas for which $\bm{r}(m)$ values are not significantly large. 
\end{remark}

\subsection{Modified Geometric Decoder}
\label{ssec:Modified Geometric Decoder}
Geometric decoding in~\autoref{ssec:GCD Decoding} involves utilizing the physical positions of the antennas to decode the DoA at each antenna. In~\autoref{alg:Geometric}, the critical step involved is quantizing the signal $\mathbf{y}(m)$. If the quantization is in error, further steps lead to wrong $\theta_m$ characterization. In this type of decoding, we increase the quantization size i.e., several neighboring points of ${t_m}$ which is a quantized point of $\mathbf{y}(m)$ are considered for further steps of decoding. This modified step of decoding helps to better quantize the signal among those antennas that suffer noisy channel conditions. The~\autoref{alg:Geometric modified} presents the increased quantization interval size based, specifically modified geometric decoder.

\begin{algorithm}

\caption{Modified geometric decoder}
\begin{algorithmic}[1] 
\State $L \gets \emptyset $
\State $\bm{r}(m) \gets \text{gcd}(\mathcal{A},N)$
\For{$m=1$ to $M$}
\If{$\bm{r}(m)=1$}
    \State $t_m \gets \lfloor (\arg(\mathbf{y}(m)) *(N/2\pi) + 1/2)\rfloor$ 
    \For{$i=t_m-k$ to $t_m+k$}
        \State ${\theta^i_m} \gets (\mathcal{A}(m)^{-1}*i) \text{ mod } N$
        \State $L \gets \{L, {\theta^i_m}\}$    
    \EndFor
    
\Else
    \State $t_m \gets \lfloor (\arg(\mathbf{y}(m)) *(N/(\bm{r}(m) *2\pi)) + 1/2)\rfloor$ 
    \State $\bm{t_m'} \gets \text{mod}((0:N-1)*d_m/\bm{r}(m), N/\bm{r}(m))$
    \For{$i=t_m-k$ to $t_m+k$}
        \State $\bm{\theta^i_m} \gets \text{find}(i=\bm{t'_m})$
        \State $L \gets \{L, \bm{\theta^i_m}\}$
    \EndFor
    
\EndIf
\EndFor
\State $\hat{\theta}  \gets \text{mode}(L)$

\end{algorithmic}
\label{alg:Geometric modified}
\end{algorithm}
\vspace{-2mm}
Step 1-5, 11, and 12 remain same as~\autoref{alg:Geometric}. However in step 6-8 and 13-15, the modified geometric decoder now considers $k$ neighboring points on unit circle for the quantization to further decode $\theta^i_m$. The superscript $i$ denotes the DoA for any quantization point in $[t_m-k, t_m+k]$ interval. Step 19 will declare the most repeated $\theta_m$ across all entries of quantization points and antennas as the decoded $\theta$.
\begin{remark}
\label{rmk: Modified geometric}
    Note that size of the set in~\autoref{alg:Geometric modified} is larger than~\autoref{alg:Geometric} considering the fact we are counting extra $2k$ points for decoding which are neighboring $t_m$. The overall complexity can be limited to $\mathcal{O}(M^2)$ ignoring the constant factors. However the final run-time complexity depends on the receiver choosing the value of $k$ for interval size and choosing a subset of antennas that have lower $\bm{r}(m)$ values. 
\end{remark}

\subsection{Geometric-reduced MAP}
\label{ssec:Geometric-reduced MAP}
The geometric decoder and modified geometric decoder considered the most repeated $\theta_m$ that occurs at each antenna. However even though we consider all possibilities of $\theta_m$ in-order to declare DoA $\theta$ by considering the most number of repetitions, the candidate $\theta$ is not guaranteed to be the most repeated one. The final decision step of declaring the candidate $\theta$ on most number of repetitions could go wrong when other values $\theta_m$ dominate across $[M]$. 
In this type of decoder, we consider top $g$ repetitions in set $L$ of the modified geometric decoder as presented in~\autoref{alg:Geometric-reduced MAP}. 

\begin{algorithm}
\caption{Geometric-reduced MAP}
\begin{algorithmic}[1] 
\State $L\gets \emptyset$
\State $\bm{r}(m) \gets \text{gcd}(\mathcal{A},N)$
\If{$\bm{r}(m)=1$}
    \State $t_m \gets \lfloor (\arg(\mathbf{y}(m)) *(N/2\pi) + 1/2)\rfloor$ 
    \For{$i=t_m-k$ to $t_m+k$}
        \State ${\theta^i_m} \gets (\mathcal{A}(m)^{-1}*i) \text{ mod } N$
        \State $L \gets \{L, {\theta^i_m}\}$    
    \EndFor
    
\Else
    \State $t_m \gets \lfloor (\arg(\mathbf{y}(m)) *(N/(\bm{r}(m) *2\pi)) + 1/2)\rfloor$ 
    \State $\bm{t_m'} \gets \text{mod}((0:N-1)*d_m/\bm{r}(m), N/\bm{r}(m))$
    \For{$i=t_m-k$ to $t_m+k$}
        \State $\bm{\theta^i_m} \gets \text{find}(i=\bm{t'_m})$
        \State $L \gets \{L, \bm{\theta^i_m}\}$
    \EndFor
    
\EndIf
\State Define the set ${S}$ that gets $\text{top-}g$ indices of {L}.  
\State $\hat{\theta} = \argmax\limits_{s \in S} |\mathbf{y}^{\rm H} * \mathcal{C^{BC}}[:, s]|$

\end{algorithmic}
\label{alg:Geometric-reduced MAP}
\end{algorithm}
\vspace{-2mm}
Until step 16, the working of geometric-reduced MAP is same as the~\autoref{alg:Geometric modified}. However, we consider set $S$ as top $g$ repeated $\theta_m$ values in set $L$ at step 17 and then match the received signal $\mathbf{y}(m)$ with corresponding codewords in the codebook $\mathcal{C^{BC}}$ in step 18 to declare the candidate codeword uniquely corresponding to the DoA $\theta$.

\begin{remark}
    The overall complexity can be limited to $\mathcal{O}(M^2)$, up to constant multiplicative factor. However depending on $k$ and the subset of antennas chosen by the receiver as explained in~\autoref{rmk: Modified geometric}, with the value of $g$ in geometric-reduced MAP decoder, the constant multiplicative factor can be substantially high depending on implementation details. We identify that this new decoding algorithm is better than any previous  geometric decoders.   
\end{remark}
In~\autoref{sec:Simulation}, we present the performance of geometry based decoders along with other decoders mentioned before via simulations in MATLAB. Before proceeding to next section, let us analyze more on the code structure and decoders ability to decode any single source signal arrival.
\subsection{Subspace Distance of Bose-Chowla Code}
\label{ssec:Subspace Bose-Chowla}
The principal advantage of using Bose-Chowla sensing subspace codes is to ensure that any DoA $\theta$ impinges uniquely on all $M$ antennas with minimal interference. Such an array geometry design ensures all codewords to be nearly orthogonal with $d^{(s)}_\text{min}(\mathcal{C^{(BC)})}$ approaching unity for all configurations of $M$~\cite{Mahdavifar_SS}. The [\cite{Mahdavifar_SS},Theorem~2] provides a lower bound on $d^{(s)}_\text{min}(\mathcal{C^{(BC)})}$ which is almost as close to unity for values of $M>10$. Let us now examine the upper bound for $d^{(s)}_\text{min}(\mathcal{C^{(BC)}})$. Recall from~\cite{Mahdavifar_SS} that, subspace distance between any two codewords $\mathbf{c}(\alpha),\mathbf{c}(\alpha') \in \mathcal{C^{(BC)}}$ is given by,
\begin{align}
\label{eqn:Bose-Chowla Subspace}
    d^{(s)}(\mathbf{c}(\alpha),\mathbf{c}(\alpha')) = 1 - \frac{1}{M^2}\left|\sum_{m=1}^M(\alpha^*\alpha')^{d_m}\right|^2
\end{align}
where $.^*$ denotes complex conjugation. We can now define minimum subspace distance from~\eqref{eqn: min subspace distance} as,
\begin{align}
\label{eqn:Bose-Chowla min dist}
    d^{(s)}_{\text{min}}(\mathcal{C^{(BC)}}) \overset{\text{def}}{=} \min_{\alpha,\alpha' \in \zeta(N),\alpha\neq\alpha}  d^{(s)}(\mathbf{c}(\alpha),\mathbf{c}(\alpha')).
\end{align}
Given~\eqref{eqn:Bose-Chowla Subspace}, maximizing~\eqref{eqn:Bose-Chowla min dist} is equivalent to minimizing (with respect to $\{d_m\}_{m=1}^M$),
\begin{align}
\label{eqn:Simplified alpha}
    \max_{\alpha\neq1, \alpha\in \zeta(N)} \left|\sum_{m=1}^M \alpha^{d_m} \right|^2
\end{align}
since $\alpha'\neq\alpha$, $\alpha^*\alpha' \in \zeta(N)$ and $\alpha^*\alpha' \neq 1$ for $\alpha', \alpha \in \zeta(N)$. Now the above quantity in~\eqref{eqn:Simplified alpha} can be written as,
\begin{align}
    \left|\sum_{m=1}^M \alpha^{d_m} \right|^2 &=   \left(\sum_{m=1}^M \alpha^{d_m} \right) \left(\sum_{m=1}^M \alpha^{d_m} \right)^* \\
    &= M + \sum_{\mathclap{a,b \in [M], a\neq b}} \alpha^{d_a-d_b}.
\end{align}
Since $\alpha\in \zeta(N)$, we have,
\begin{align}
   \sum_{\mathclap{a,b \in [M], a\neq b}} \alpha^{d_a-d_b} = \sum_{\mathclap{q \in \mathcal{A}}}\alpha^q &= \hspace{10pt} -\sum_{\mathclap{q' \in [N-1]\cup\{0\}\setminus\mathcal{A}^-}} \alpha^{q'}\\
   \left| \sum\limits_{\substack{a,b \in [M], \\ a\neq b}} \alpha^{d_a - d_b} \right| &\geq M + 1.
\end{align}
This is because there exists at-least one element in the set $\{[N-1]\cup\{0\}\setminus\mathcal{A}^-\}$. Therefore from~\eqref{eqn:Bose-Chowla Subspace},
\begin{align}
\label{eqn: Upperbound}
    d^{(s)}_\text{min}(\mathcal{C^{(BC)}}) < 1 - \frac{1}{M}.
\end{align}
Now that the upper bound is available via~\eqref{eqn: Upperbound}, we can combine the result for $d^{(s)}_\text{min}(\mathcal{C^{(BC)}})$ in~\cite{Mahdavifar_SS} together as,
\begin{align}
   1 - \frac{2}{M} < d^{(s)}_\text{min}(\mathcal{C^{(BC)}}) < 1 - \frac{1}{M}.
\end{align}

\begin{remark}
 Closely evaluating minimum distance with lower and upper bounds, we can see that $\forall\mathbf{c}(\alpha)\in \mathcal{C^{(BC)},\alpha\in\zeta(N)}$, the codewords, $\mathbf{c}(\alpha)$ are well spaced apart with almost no closeness leading to perfect orthogonal subspaces. This demonstrates the difficulty in designing more efficient decoders, as when the decoder fails, all other codewords become almost equally likely rendering efforts to implement an iterative decoder, as in classical block codes, inapplicable. Therefore, it is possible that a linear-time decoder, i.e., a decoder with complexity $\mathcal{O}(M)$, with reasonable performance may not exist for the considered sensing subspace codes. This opens a new avenue of research to design codes that may not be optimal from a minimum distance design perspective, yet they offer efficient linear-time decoders.
\end{remark}

\section{Simulation Results}
\label{sec:Simulation}
We simulate the proposed new decoders for Bose-Chowla sensing subspace codes
in MATLAB with $M=19$. 
The difference set $\mathcal{A}$ is calculated using SageMath. 

\subsection{Window Decoders}
\label{ssec: Window Decoder simulations}
\begin{figure}[t]
\centering
\captionsetup{font=scriptsize, labelfont=scriptsize}
\captionsetup[subfigure]{font=scriptsize, labelfont=scriptsize}
\subcaptionbox{Window decoders.\label{Res: Window Decoding}}{%
{\resizebox{4cm}{!}{\begin{tikzpicture}
\definecolor{black}{rgb}{0, 0.0, 0}%
\definecolor{blue}{rgb}{0, 0, 1}%
\definecolor{red}{rgb}{1, 0, 0}%



\begin{semilogyaxis}[
font=\footnotesize,
width=7cm,
height=5cm,
scale only axis,
xmin=-10,
xmax=30,
xtick = {-10,0,..., 30},
xlabel={\huge {SNR}},  
ymin=0.0001,
ymax=1.2,
ytick={0.0001, 0.001, 0.01, 0.1, 1.2},
yticklabels={$10^{-4}$, $10^{-3}$, $10^{-2}$, $10^{-1}$, $1$},
ylabel={\huge ${P_e}$},
ylabel near ticks,
legend style={
    font=\LARGE,           
    at={(0.5,-0.35)},              
    anchor=north,
    draw=none,
    fill=white,
    legend cell align=left,
    /tikz/every even column/.append style={column sep=0.0cm}  
},
legend columns = 1,                
xticklabel style={font=\huge},
yticklabel style={font=\huge},
]


\addplot[color=red, line width=2.0pt, mark=none]table[row sep=crcr]{%
-10.0   1\\
-7.75   1\\
-5.5    1\\
-3.25   1\\
-1.0    1\\
1.25    1\\
3.5     0.6516\\
5.75    0.009\\
8.0     1e-6\\
10.25	1e-12\\
12.5	0\\
14.75	0\\
17.0	0\\
19.25	0\\
21.5	0\\
23.75	0\\
26.0	0\\
28.25	0\\
30.5	0\\
32.75	0\\
35.0    0\\
};
\addlegendentry{{Upper Bound}}

\addplot [color=blue, solid, dashed, mark= o, mark options = {solid}, line width=2.0pt]
  table[row sep=crcr]{%
-10.0	0.926\\
-7.75	0.852\\
-5.5	0.628\\
-3.25	0.312\\
-1.0	0.069\\
1.25	0.001\\
3.5	    1e-6\\
5.75	0\\
8.0	    0\\
10.25	0\\
12.5	0\\
14.75	0\\
17.0	0\\
19.25	0\\
21.5	0\\
23.75	0\\
26.0	0\\
28.25	0\\
30.5	0\\
32.75	0\\
35.0    0\\
};
\addlegendentry{{MAP}}


\addplot [color=black, solid, dashed, mark= triangle, mark options = {solid}, line width=2.0pt]
  table[row sep=crcr]{%
-10.0    0.966\\
-7.75   0.943\\
-5.5    0.854\\
-3.25   0.684\\
-1.0    0.396\\
1.25    0.144\\
3.5     0.020\\
5.75    1e-6\\
8.0     0\\
10.25	0\\
12.5	0\\
14.75	0\\
17.0	0\\
19.25	0\\
21.5	0\\
23.75	0\\
26.0	0\\
28.25	0\\
30.5	0\\
32.75	0\\
35.0    0\\
};
\addlegendentry{{2-Win}}

\addplot [color=black, solid, dashed, mark= diamond, mark options = {solid}, line width=2.0pt]
  table[row sep=crcr]{%
-10.0   0.971\\
-7.75   0.946\\
-5.5    0.893\\
-3.25   0.769\\
-1.0    0.596\\
1.25    0.369\\
3.5     0.156\\
5.75    0.044\\
8.0     0.002\\
10.25	1e-6\\
12.5	0\\
14.75	0\\
17.0	0\\
19.25	0\\
21.5	0\\
23.75	0\\
26.0	0\\
28.25	0\\
30.5	0\\
32.75	0\\
35.0    0\\
};
\addlegendentry{{3-Win}}


\addplot [color=black, solid, dashed, mark= x, mark options = {solid}, line width=2.0pt]
  table[row sep=crcr]{%
-10     0.969\\
-7.75   0.959\\
-5.5    0.925\\
-3.25   0.843\\
-1.0    0.729\\
1.25    0.61\\
3.5     0.453\\
5.75    0.333\\
8.0     0.263\\
10.25   0.217\\
12.5    0.14\\
14.75   0.087\\
17.0    0.063\\
19.25   0.031\\
21.5    0.008\\
23.75   0.004\\
26.0    1e-6\\
28.25   0.0\\
30.5    0.0\\
32.75   0.0\\
35.0    0.0\\
};
\addlegendentry{{5-Win}}


\end{semilogyaxis}

\end{tikzpicture}%
}}
}
\hspace{0.0cm}
\subcaptionbox{Array geometry based decoders.\label{Res: GCD}}{%
{\resizebox{4cm}{!}{\begin{tikzpicture}
\definecolor{black}{rgb}{0, 0.0, 0}%
\definecolor{blue}{rgb}{0, 0, 1}%
\definecolor{red}{rgb}{1, 0, 0}%



\begin{semilogyaxis}[
font=\footnotesize,
width=7cm,
height=5cm,
scale only axis,
xmin=-10,
xmax=30,
xtick = {-10,0,..., 30},
xlabel={\huge {SNR}},  
ymin=0.0001,
ymax=1.2,
ytick={0.0001, 0.001, 0.01, 0.1, 1.2},
yticklabels={$10^{-4}$, $10^{-3}$, $10^{-2}$, $10^{-1}$, $1$},
ylabel={\huge ${P_e}$},
ylabel near ticks,
legend style={
    font=\LARGE,           
    at={(0.5,-0.35)},             
    anchor=north,
    draw=none,
    fill=white,
    legend cell align=left,
    /tikz/every even column/.append style={column sep=0.0cm}   
},
legend columns = 1,                
xticklabel style={font=\huge},
yticklabel style={font=\huge},
]


\addplot[color=red, line width=2.0pt, mark=none]table[row sep=crcr]{%
-10.0   1\\
-7.75   1\\
-5.5    1\\
-3.25   1\\
-1.0    1\\
1.25    1\\
3.5     0.6516\\
5.75    0.009\\
8.0     1e-6\\
10.25	1e-12\\
12.5	0\\
14.75	0\\
17.0	0\\
19.25	0\\
21.5	0\\
23.75	0\\
26.0	0\\
28.25	0\\
30.5	0\\
32.75	0\\
35.0    0\\
};
\addlegendentry{{ Upper Bound}}

\addplot [color=blue, solid, dashed, mark= o, mark options = {solid}, line width=2.0pt]
  table[row sep=crcr]{%
-10.0	0.940000000000000\\
-7.75	0.821000000000000\\
-5.5	0.621000000000000\\
-3.25	0.309000000000000\\
-1.0	0.0700000000000000\\
1.25	0.00200000000000000\\
3.5	    1e-6\\
5.75	0\\
8.0	    0\\
10.25	0\\
12.5	0\\
14.75	0\\
17.0	0\\
19.25	0\\
21.5	0\\
23.75	0\\
26.0	0\\
28.25	0\\
30.5	0\\
32.75	0\\
35.0    0\\
};
\addlegendentry{{ MAP}}


\addplot [color=black, solid, dashed, mark= triangle, mark options = {solid}, line width=2.0pt]
  table[row sep=crcr]{%
-10.0    0.904000000000000\\
-7.75   0.772000000000000\\
-5.5    0.573000000000000\\
-3.25   0.256000000000000\\
-1.0    0.0650000000000000\\
1.25    0.00900000000000000\\
3.5     0.00200000000000000\\
5.75    1e-6\\
8.0     0\\
10.25	0\\
12.5	0\\
14.75	0\\
17.0	0\\
19.25	0\\
21.5	0\\
23.75	0\\
26.0	0\\
28.25	0\\
30.5	0\\
32.75	0\\
35.0    0\\
};
\addlegendentry{{ Geometric-reduced MAP}}

\addplot [color=black, solid, dashed, mark= diamond, mark options = {solid}, line width=2.0pt]
  table[row sep=crcr]{%
-10.0   0.974000000000000\\
-7.75   0.947000000000000\\
-5.5    0.927000000000000\\
-3.25   0.838000000000000\\
-1.0    0.705000000000000\\
1.25    0.567000000000000\\
3.5     0.312000000000000\\
5.75    0.142000000000000\\
8.0     0.0290000000000000\\
10.25	0.00200000000000000\\
12.5	1e-6\\
14.75	0\\
17.0	0\\
19.25	0\\
21.5	0\\
23.75	0\\
26.0	0\\
28.25	0\\
30.5	0\\
32.75	0\\
35.0    0\\
};
\addlegendentry{{ Modified Geometric}}


\addplot [color=black, solid, dashed, mark= x, mark options = {solid}, line width=2.0pt]
  table[row sep=crcr]{%
-10     0.986000000000000\\
-7.75   0.994000000000000\\
-5.5    0.977000000000000\\
-3.25   0.971000000000000\\
-1.0    0.953000000000000\\
1.25    0.916000000000000\\
3.5     0.840000000000000\\
5.75    0.751000000000000\\
8.0     0.640000000000000\\
10.25   0.500000000000000\\
12.5   0.371000000000000\\
14.75   0.218000000000000\\
17.0    0.133000000000000\\
19.25   0.0720000000000000\\
21.5    0.0220000000000000\\
23.75   0.00600000000000000\\
26.0    0.00200000000000000\\
28.25   1e-6\\
30.5    0.0\\
32.75   0.0\\
35.0    0.0\\
};
\addlegendentry{{ Geometric}}


\end{semilogyaxis}

\end{tikzpicture}%
}}
}
\caption{Empirical probability of error of window and array geometry based decoders for BCG code vs SNR (in dB scale) for $M =19$.}
\label{fig:main}
\vspace{-5mm}
\end{figure}
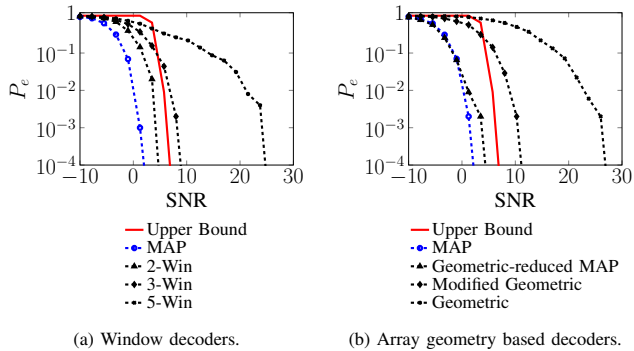

The \textit{window decoder} with 
sizes, $z =2, 3, 5$ is evaluated for the decoder performance as shown in~\autoref{Res: Window Decoding}. The proposed decoder for window size $z=2$ is within the upper bound for the probability of error
~\cite{Mahdavifar_SS}. However for the window sizes $z=3$  and $5$, the probability of error is much higher 
because, the noise 
across each $m^{th}$ antenna could correlate with combined codewords, misleading the decoder to a wrong window to a wrong codeword. There exists a tradeoff with window size $z$ and a given SNR value required for a particular value of $M$ to guarantee a required error performance.

\subsection{Geometric Decoders}
\label{ssec:Geometric Decoding Sim}
The array geometry based decoders are evaluated for their performance as mentioned in~\autoref{ssec:GCD Decoding},~\autoref{ssec:Modified Geometric Decoder} and~\autoref{ssec:Geometric-reduced MAP} $\forall m \in [M]$ as shown in~\autoref{Res: GCD}. The \textit{geometric decoder} performs the least among all of the geometry based decoders. The time complexity is $\mathcal{O}(M^2)$, up to constant factors. The \textit{modified geometric decoder} is better than geometric decoder for a given SNR value and $k=9$. The size of the set grows at the sub-cubic storage complexity along with the total run-time complexity of the algorithm. In \textit{geometric-reduced MAP}, instead of considering single most repeated $\theta$ among $M$ antennas, we consider top-$g$ repeated $\theta$ values to perform MAP among those $g$ codewords. The $g$ value chosen is $N/2$, almost matching the received signal with the half the number of total codewords that are most repeated. As observed in~\autoref{Res: GCD}, the performance smoothly approaches MAP decoder with run-time complexity sub-cubic order of $M$. 
\section{Conclusion}
In our proposed work, we present novel efficient decoding algorithms, mainly two categories-\textit{window} and \textit{geometry} based algorithms that efficiently estimate the DoA in sparse antenna array systems. This new set of decoders provide flexibility in run-time complexity with many parameters at the receiver to choose, offering smooth tradeoff between error performance and time complexity. 
The proposed geometric-reduced MAP decoder smoothly approaches MAP decoder performance for cubic complexity of number of antennas.
\label{sec:Conclusion}

\bibliographystyle{IEEEtran}
\bibliography{ref}

\end{document}